\documentclass[twocolumn,showpacs,preprintnumbers,amsmath,amssymb]{revtex4-1}
\usepackage{amssymb}					
\usepackage{amsthm}					
\usepackage{newlfont}					
\usepackage{graphicx}					
\usepackage{longtable,rotating}			
\usepackage{graphicx}
\usepackage{dcolumn}
\usepackage{bm}
\usepackage{amsmath}%
\usepackage{amsfonts}%

\usepackage{t1enc}
\usepackage[utf8]{inputenc}
\usepackage{ascii}

\usepackage{color}
\newcommand{\vectorr}[1]{\mathbf{#1}}                                                        

\newcommand{\pgftextcircled}[1]{                                                                    
    \setbox0=\hbox{#1}%
    \dimen0\wd0%
    \divide\dimen0 by 2%
    \begin{tikzpicture}[baseline=(a.base)]%
        \useasboundingbox (-\the\dimen0,0pt) rectangle (\the\dimen0,1pt);
        \node[circle,draw,outer sep=0pt,inner sep=0.1ex] (a) {#1};
    \end{tikzpicture}
}

\newcommand{\tra}[1]{\textnormal{trace} \: #1}
\newcommand{\sign}[1]{\textnormal{sign\,}#1}

\newcounter{proofcount}

\begin{document}

\title{Subcritical Turing bifurcation and the morphogenesis of localised
patterns}

\author{V\'ictor Bre\~{n}a--Medina$^{1,2}$ \& Alan Champneys$^2$}
\affiliation{$^1$Departamento de Nanotecnolog\'ia, Centro de F\'isica Aplicada y Tecnolog\'ia Avanzada, Universidad Nacional Aut\'onoma de M\'exico, Juriquilla No. 3001, Quer\'etaro 76230, M\'exico \\ 
$^2$Department of Engineering Mathematics, University of Bristol, Queen's Building, University Walk, Bristol BS8 1TR, United Kingdom}
\date{\today}

\begin{abstract}
Subcritical Turing bifurcations of reaction-diffusion systems in
large domains lead to spontaneous onset of well-developed localised
patterns via the homoclinic snaking mechanism. This phenomenon is shown
to occur naturally when balancing source and loss effects are included
in a typical reaction-diffusion system, leading to a 
super/subcritical transition.  Implications are
discussed for a range of physical problems, arguing that
subcriticality leads to naturally robust phase transitions to
localised patterns.
\end{abstract}

\maketitle

\section{Introduction}

Reaction-diffusion systems are known to give rise to a wide variety of
stationary and oscillatory patterns, see~e.g.~\cite{murra02,vanag,kondo01}. 
The primary mechanisms for explaining
transition from quiescent to  patterned states
is the instability first described by Alan Turing \cite{turing}. 
Such patterns are used to explain
diverse physical phenomena, such as gas
discharge dynamics \cite{purwens}, active fluid behaviour \cite{bois}
and tumour growth~\cite{khain}. Now diffusion-driven instability, or
{\em Turing bifurcation}, 
is a key part of any graduate course on nonlinear far-from equilibrium 
physics or biology. For systems in large
domains however, many different wave numbers can become unstable in Turing
bifurcations for nearby parameter values 
and mode interactions can lead to a remarkable richness
in patterns and their dynamics, see e.g.~\cite{Hoyle}.

A different explanation of {\em localised} pattern formation has emerged in
recent years; the so-called homoclinic snaking mechanism 
\cite{woods,beck}. 
The 1D generalised Swift--Hohenberg equation with competing nonlinear terms
is a canonical model for such analysis \cite{burke01,burke02}. 
In 2D a richness of localised
stripy, spotty, hexagonal, square-wave and target-like patterns have
been observed \cite{Daniele,lloyd1,lloyd2}. The mechanism has been shown to
underlie many physical observations such as 
the onset of turbulent spots in
plain Couette flow~\cite{schneider}, stationary patterns in 
binary convection \cite{knobloch} and localised modes in optical
cavities \cite{gomila}.

One of the distinctions between the homoclinic snaking and Turing
bifurcation pattern formation theories is that the Swift--Hohenberg 
equation has variational structure, which can be linked to the free
energy of the system. General systems of reaction 
diffusion equations for which the Turing mechanism applies do not
typically have such variational structure. 
However, the snaking mechanism still applies to 
Swift--Hohenberg equations with broken variational structure 
\cite{Houghton}, provided spatial reversibility is retained, albeit 
stationary asymmetric patterns are lost.

The purpose of this paper then is to show how the connection
between homoclinic snaking and Turing instability analysis gives 
a robust explanation for the
formation of localised patterns in reaction-diffusion systems. 
We show that this robustness arises from inclusion of source
and loss terms in reaction-diffusion models, which realistic effects
are often ignored in canonical models. For example, a model
equivalent to the system we study below but without source and loss
terms gives rise to wave pinning but no localised patterns
\cite{mori}.
Inclusion of such terms
naturally breaks material conservation, paves the way for effective
competing nonlinear terms and in turn this allows Turing bifurcations
to become {\em sub-critical}. 
We show that this subcriticality is
equivalent to the key condition for homoclinic snaking to occur in
long domains (see also~\cite{Daniele}).  Hence, upon considering long
domains, we can set the backbone conditions under which
reaction-diffusion systems naturally give rise to spots and pulses,
rather to than just spatially extended patterns (see also~\cite{vanag}
for further experimental and theoretical evidence).

It is worth mentioning earlier related work of 
Yochelis {\em et al.}~\cite{yochelis02} who performed numerical bifurcation analysis on the Gierer--Meinhardt system, which 
includes a rational nonlinearity. They also found the
existence of a bistability region and a subcritical Turing bifurcation that leads to 
homoclinic snaking; see also~\cite{yochelis01} for further details. 
The novelty of the present paper is to show that such scenarios are in some sense generic and can occur for a pure-power nonlinear system, given
the presence of source and loss terms.

\section{Local analysis}

For the ease of
explanation, we shall perform our detailed calculations in 1D in space. 
Extension to higher spatial
dimension is in principle possible as we shall indicate in what follows,
although there are additional considerations due to the range of different
spatial symmetries of underlying patterns, stripes, rolls, hexagonal lattices,
among others, see \cite{lloyd1,lloyd2,lloyd3}.  
We shall also apply the theory to reaction-diffusion systems with
just two interacting species and a single nonlinear interaction term. 
Application to more complex systems is in principle straightforward, because
the theory is built upon the principle of normal-form reduction. 

Consider a reaction-diffusion system 
\begin{subequations}\label{eq:turing}
	\begin{gather}
		U_t = D_1 U_{xx} + F(U,V;\mu)\,, \label{eq:turinga}\\
	        V_t = D_2 V_{xx} + G(U,V;\mu)\,, \label{eq:turingb}
	\end{gather}
\end{subequations}
for $x\in (-L/2,L/2)$, subject to homogeneous Neumann boundary conditions.
Here, parameters $D_1$ and $D_2$ are diffusion coefficients 
and $F$ and $G$ are sufficiently smooth functions.
Without loss of generality timescales have been scaled to unity, but
the length scale $L$ is retained. 

The linear analysis provides of the usual conditions under which Turing bifurcations occur, see
\cite{murra02}. Thus, suppose there exists an isolated homogeneous equilibrium $(U,V)^T=(U_0,V_0)^T$.
Upon substituting the incremental variables $U=U_0+u$ and $V=V_0+v$
into system \eqref{eq:turing}, we obtain 
\begin{equation}
\begin{pmatrix} 
		u_t \\ 
                v_t \end{pmatrix} 
= \begin{pmatrix} D_1 u_{xx} \\ 
D_2 v_{xx}  \end{pmatrix}
+ \vectorr A
(\mu) \begin{pmatrix} u \\ v \end{pmatrix}  
+
\begin{pmatrix} 
f(u,v;\mu) \\
g(u,v;\mu)
\end{pmatrix} 
\label{eq:linturing}
\end{equation}
where $\vectorr A = \{ a_{ij} \}$ with 
$a_{11}(\mu)=F_U$, $a_{12}(\mu)=F_V$, $a_{21}(\mu)=G_U$, $a_{22}(\mu)=G_V$ evaluated
at the steady state and $f$ and $g$ gather all remaining higher-order
terms. Under the usual assumptions of Turing bifurcation analysis, we first need to assume that the homogeneous steady
state is stable in the absence of diffusion; that is
$\tra{\vectorr A}<0$ and $\det{\vectorr A}>0$.

To find diffusion-driven instability with spatial wavenumber $\kappa$ we
look for modes of the form
\begin{equation}
\cos\left(\kappa (x -L/2)\right)\begin{pmatrix} C_1\\ C_2 
\end{pmatrix}, \quad C_1^2+C_2^2>0\,,
\label{eq:modek}
\end{equation} 
where $\kappa=\kappa_m=m\pi/L$ for some positive integer mode number $m$. This
leads to the search for zero eigenvalues $\lambda$ of the 
matrix 
\begin{gather}\label{eq:eigenA}
\vectorr A_\kappa(\mu) = \begin{pmatrix}  a_{11} - D_1 \kappa^2 & a_{12} \\
                       a_{21} & a_{22}-D_2 \kappa^2
\end{pmatrix},
\end{gather}
with wavenumber $\kappa=\sqrt{(D_2 a_{11} + D_1
  a_{22})/(D_1D_1)}$. The condition for instability is that $\kappa$
is real, which is guaranteed for a stable equilibrium (with $\sign
\left(a_{11}a_{22} \right)<0$) if $D_1D_2 \ll \max \left\{ D_1^2,
D_2^2\right \}$.  Without loss of generality, suppose $a_{22}<0$ and
that $D_1 \ll D_2$ which, in comparison to the $V$ component, implies
that the $U$ component diffuses slowly.

\begin{figure}
\begin{center}
(a) \hspace{3.5cm}  (b) \\
\includegraphics[width=4.4cm,height=4.4cm]{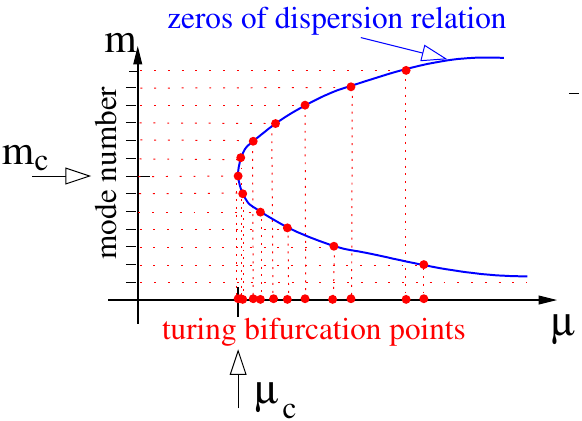} \hspace{-0.4cm}
\includegraphics[width=4.4cm,height=4.4cm]{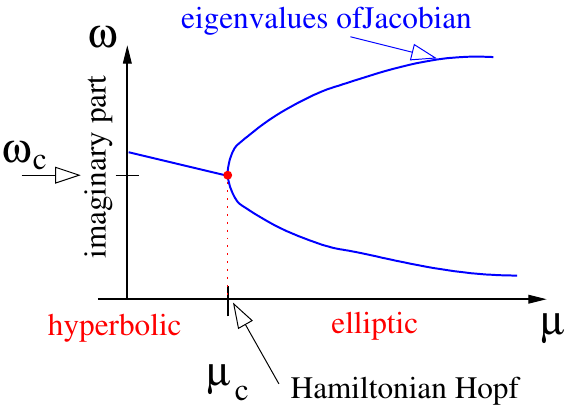}
\caption{(Color online) The equivalence between 
(a) the mode numbers accumulation point of Turing bifurcations and (b) a Hamiltonian--Hopf
bifurcation point for long domains.} 
\label{fig:turingequiv}
\end{center}
\end{figure}

We now perform a proper comparison between the two theories in question, 
in the case of a long domain $L\gg1$. We suppose that at
parameter value $\mu=\mu_c$ there is a double zero eigenvalue of
$A_\kappa$, corresponding to a large mode number $m_c$, not
necessarily an integer.  The condition for such a double root is
\begin{equation}
(D_2 a_{11} + D_1 a_{22})^2= 4 D_1 D_2 \det \vectorr A.
\label{eq:doubleroot}
\end{equation}
On the one hand, for a $L\gg 1$ there will 
be a large number of Turing bifurcations for
nearby $\kappa$-values corresponding to $\kappa=m\pi/L$ for integers
$m$ close to $m_c$.  Generically, $\mu$~will depend quadratically on
$\kappa$ close to~$\mu_c$, which would imply a double accumulation of
Turing bifurcations; one family corresponding to higher wavenumbers
$\kappa>\kappa_c$, the other to lower wavenumbers $\kappa<\kappa_c$;
see in~Fig.~\ref{fig:turingequiv}(a) the mode numbers $m$ which correspond to zeros of the dispersion relation as $\mu$ tends towards $\mu_c$.

Alternatively, in the limit $L\to\infty$, so-called spatial dynamics
can be applied (see e.g.~\cite{beck}) where steady states
$(U(x,t),V(x,t))^T=(u(x),v(x))^T$ of \eqref{eq:turing} are sought by
considering the ODE system on the real line
\begin{equation*}
D_1 u_{xx} + F(u,v; \mu)=0\:, \quad D_2 v_{xx} + G(u,v; \mu)=0\:,
\end{equation*}
as a four-dimensional dynamical system in `time' $x$. As such, the symmetry 
$(u_x,v_x)^T \to (-u_x,-v_x)^T$ and $x\to-x$ corresponds to a spatial reversibility.
In this context, a homogeneous steady state $(U_0,V_0)^T$ of the PDE corresponds
to an equilibrium $(u,u_x,v,v_x)^T=(U_0,0,V_0,0)^T$ within the fixed point set of
the reversibility. The linearisation of the system 
about such an equilibrium would take the form
\begin{equation}
\begin{pmatrix}
		u_{xx} \\ 
                v_{xx} \end{pmatrix} 
+\begin{pmatrix} a_{11}/D_1 &  a_{12}/D_1 \\ 
a_{21}/D_2 & a_{22}/D_2  \end{pmatrix} 
\begin{pmatrix} u \\ v \end{pmatrix}  = \begin{pmatrix} 0 \\ 0 \end{pmatrix}\,.
\label{eq:odejac}
\end{equation}
Such an equilibrium will undergo a transition from being hyperbolic to
elliptic at a Hamiltonian--Hopf bifurcation 
(also known as a reversible 1:1 resonance) \cite{iooss} which
occurs, under suitable non-degeneracy conditions, when there is a
double pair of complex conjugate eigenvalues $\pm i \omega$ of the
Jacobian in \eqref{eq:odejac}. Upon substituting $(u,v)^T= (A,B)^T \exp\left(i\omega  x\right)$ into \eqref{eq:odejac}, 
we find that we need a double root to
$$
D_1D_2 \omega^4 - (D_2a_{11} + D_1 a_{22}) \omega^2 + \det \vectorr A  =0\,,
$$
which leads to precisely the same condition for a 
fold Turing points with respect to $\kappa$, namely equality as in~\eqref{eq:doubleroot}; 
see Fig.~\ref{fig:turingequiv}(b).
It is straightforward to show that the condition for a criticality 
of the Turing bifurcation at the double root is precisely the 
same as the condition for the criticality of the corresponding 
Hamiltonian--Hopf; see~e.g.~\cite{burke03}. Both problems may be expressed as 
via an amplitude equation whose real part reads \cite{iooss}
\begin{equation}
Z''(\xi)  = q_1 (\mu-\mu_c)Z  +q_3 Z|Z|^2 + q_5 Z|Z|^4 \,. 
\label{eq:long}
\end{equation}

A key prediction of the homoclinic snaking mechanism~\cite{woods} is
the birth of a spatially localised mode (a homoclinic orbit in space
$x$) if the Hamiltonian--Hopf bifurcation is subcritical, $q_1q_3>0$.
For small $q_1q_3>0$, then, provided $q_1q_5<0$, an unfolding of the
normal form shows that there is a heteroclinic connection from a
background state to a non-trivial periodic orbit. Taking account of
beyond-all-orders terms in the normal form~\cite{dean,kozyreff}
enables an analysis to be undertaken in which we find infinitely many
homoclinic orbits arranged on two closed curves; see
Fig.~\ref{fig:crittrans}(b), below.

\section{Illustration for a generalised Schnakenberg system} 

In order to illustrate our findings,  we here consider the generalised Schnakenberg system, which is a spatially homogeneous form of a model proposed in \cite{payne01} of pattern formation via interaction between  active~$U$ and inactive~$V$ small G-proteins in sub-cellular-level biological 
morphogenesis:
\begin{subequations}\label{eq:fundsys}
	\begin{flalign}
	& U_t = D_1 \nabla^2 U  + k_2U^2V-(c+r)U+hV\,, \label{eq:fundsysa} \\
	& V_t = D_2 \nabla^2 V  - k_2U^2V+cU-hV+b\,,  \label{eq:fundsysb}
	\end{flalign}
\end{subequations}
in which all parameters are taken to be positive. Here the model models a non-reversible autocatalytic process, and differs from the standard Schnakenberg  and Gray--Scott systems through the presence of a production term~$b$ of the inactive
component and removal rate $r$ of the activated component. Notice that, as can easily be shown, otherwise Turing bifurcations are always supercritical in straightforward  Schnakenberg and Gray--Scott systems. 

There is a unique homogeneous equilibrium 
\begin{flalign}\label{eq:u0v0}
	U_0\equiv\frac{b}{r}\,, \qquad V_0\equiv\frac{br(c+r)}{k_2b^2+hr^2}\,.
\end{flalign}
Upon substituting the incremental variables $U=U_0+u$ 
and $V=V_0+v$ into system \eqref{eq:fundsys} in 1D, we get a system of the
form \eqref{eq:linturing}
where the coefficients of $\vectorr A$ are given by  
\begin{subequations}\label{eq:jac}
	\begin{flalign}
		& a_{11} = \frac{(c+r)\left(k_2b^2-hr^2\right)}{k_2b^2+hr^2}\,, \label{eq:jacaa}\\ 
		& a_{12} = -a_{22} = \frac{k_2b^2+hr^2}{r^2}\,, \label{eq:jaca}\\ 
		& a_{21} = \frac{chr^2-k_2b^2\left(c+2r\right)}{k_2b^2+hr^2}\,. \label{eq:jacbb}
	\end{flalign}
\end{subequations}
and the nonlinear terms
\begin{equation}\label{eq:linnonb}
	\begin{pmatrix}
	f \\
	g
	\end{pmatrix}
	\equiv
		k_2\left(u^2v+V_0u^2+2U_0uv\right)
	\begin{pmatrix}
	1 \\
	-1
	\end{pmatrix}\,.
\end{equation}
Note that the nonlinearity contains both quadratic and cubic terms
when written in these co-ordinates. It is straightforward to show 
that the steady state~$\left(U_0,V_0\right)^T$ is
asymptotically stable in the absence of diffusion provided~$c+r<8h$.

\begin{figure}
\begin{center}
(a) \hspace{3.5cm}  (b) \\
\includegraphics[width=4.4cm,height=4.4cm]{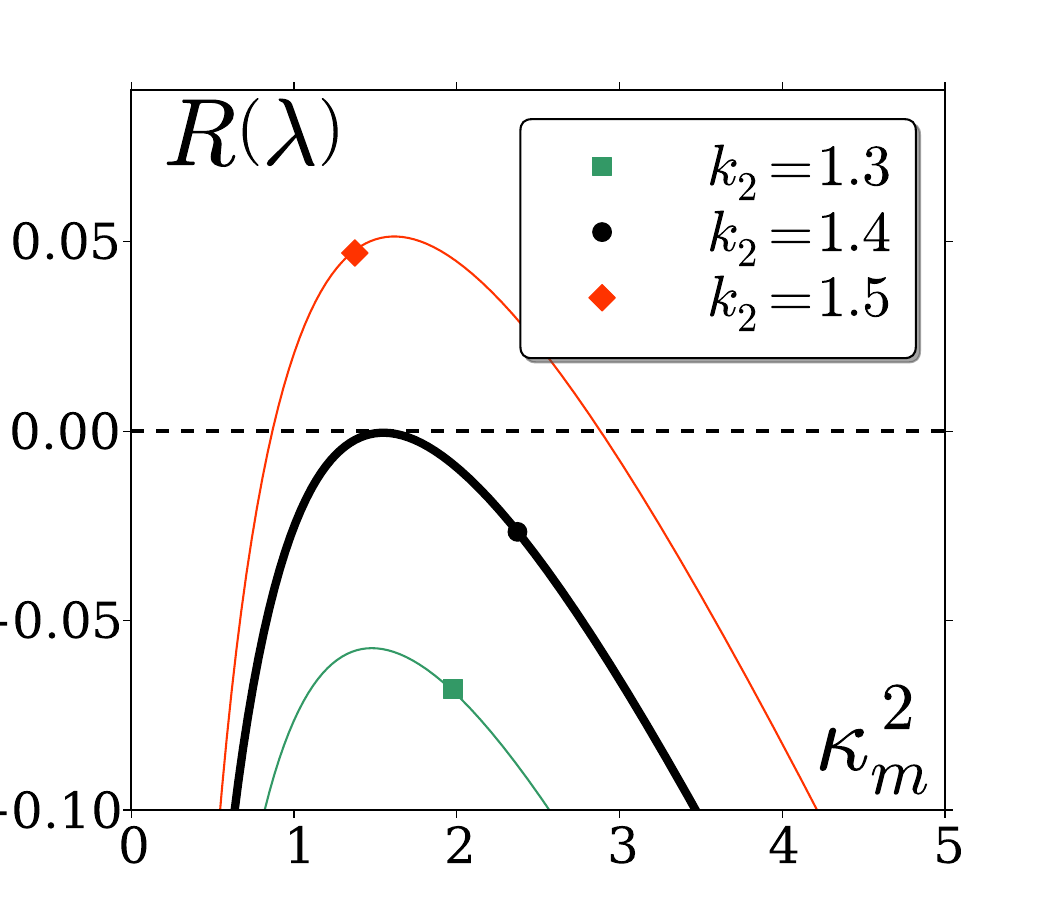} \hspace{-0.4cm}
\includegraphics[width=4.4cm,height=4.4cm]{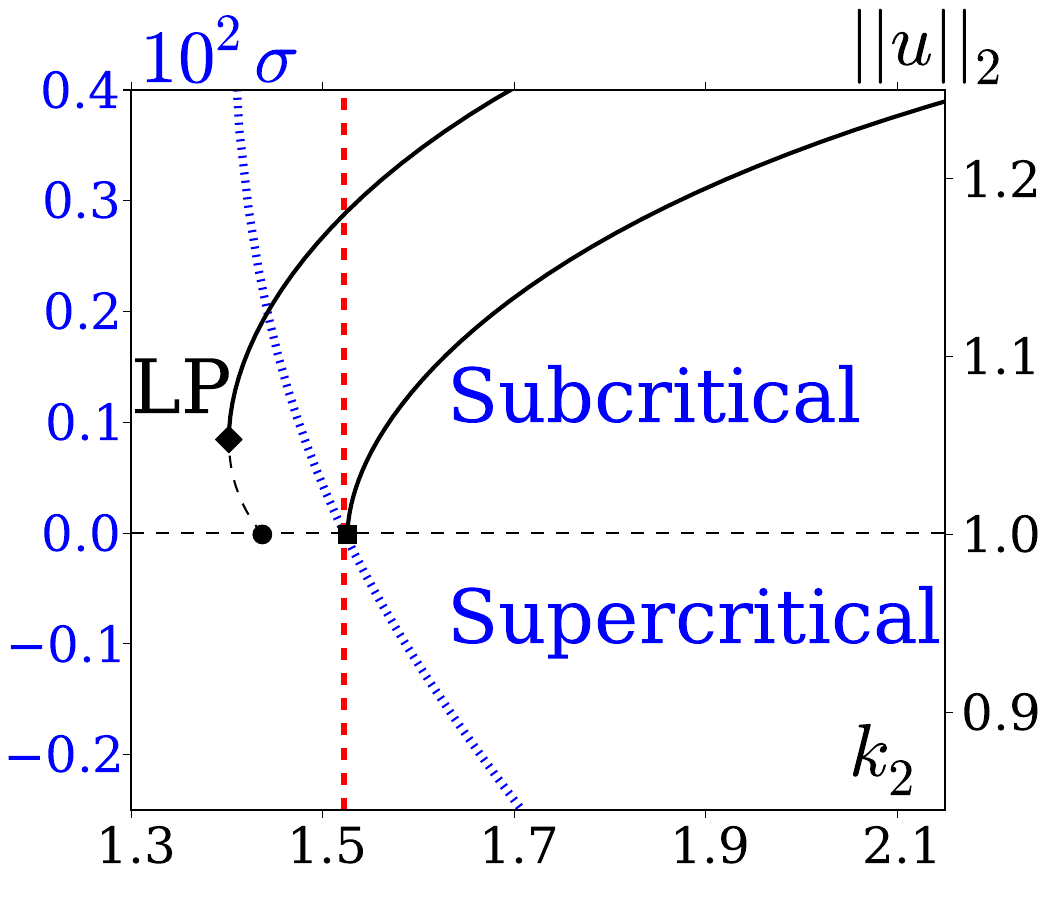}
\caption{(Color online) (a) Dispersion relation of $\vectorr A_\kappa\left(\mu\right) $ restricted to the
  eigenspace spanned by modes of the form
  \eqref{eq:modek}. The bold solid curve corresponds to where a 
 double root of~$\det\left[\vectorr A_\kappa\left(\mu\right)\right]=0$ occurs. (b)~Bifurcation diagram and
  pitchfork criticality condition; stable branches are shown as
  solid lines, the filled circle at $k_2=1.4369$ corresponds to a 
  subcritical bifurcation, and the square at $k_2=1.5258$ to a 
 supercritical bifurcation. The subcritical branch undergoes to a fold bifurcation (LP). The
  pitchfork criticality condition is depicted as a (blue) heavily dashed line, where the
  criticality transition is indicated by a (red) vertical dashed 
 line at $k_2=1.5226$. 
  }
\label{fig:Turdisper}
\end{center}
\end{figure} 
Fig.~\ref{fig:Turdisper}(a) shows the dispersion relation as
function of squared wavenumber $\kappa_m^2$, for several values of a bifurcation parameter $k_2$. 
In this and unless otherwise stated in what follows we use parameter values 
$k_2\in (0,5)$ and $b=1$, $c=1$, $r=1$, $h=1$, $D_1 =0.1$, $D_2=10$.
\begin{figure}
\begin{center}
(a) \\
\includegraphics[width=0.35\textwidth]{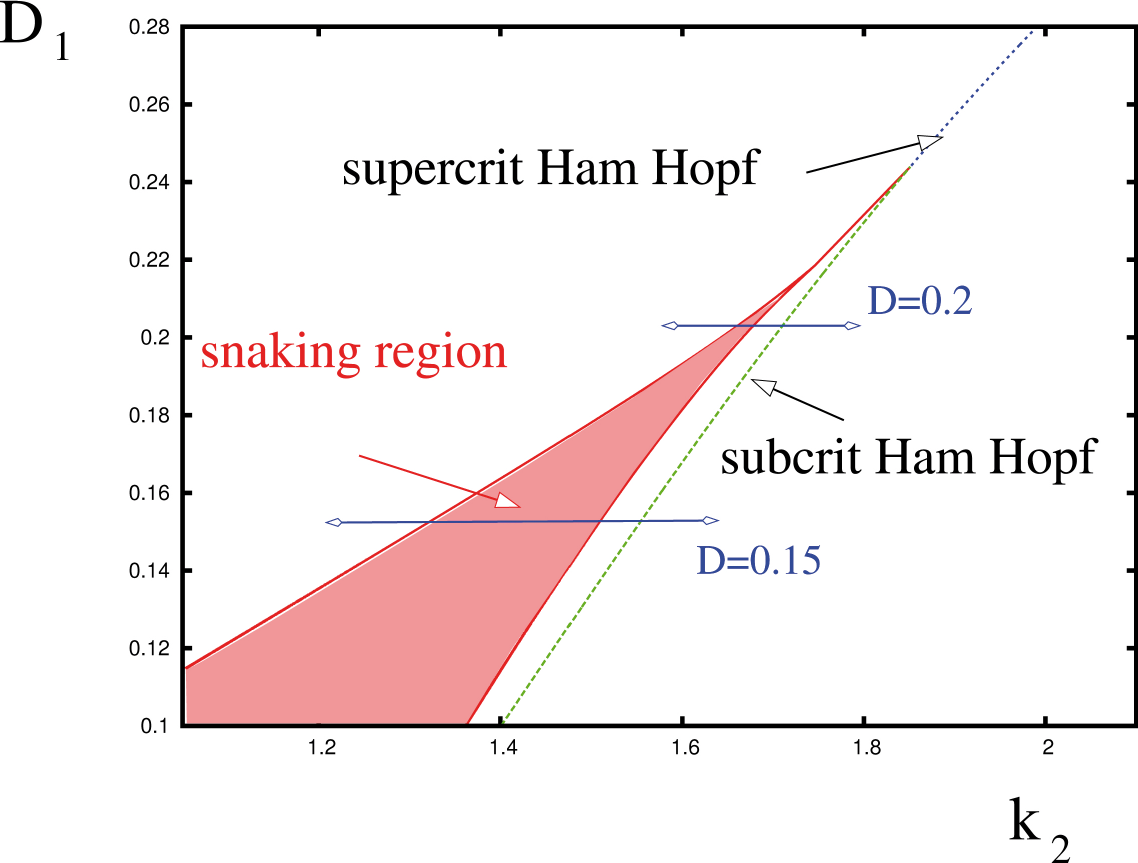}
\\
(b) \hspace{3.5cm}  (c) \\

\includegraphics[width=4.4cm,height=4.4cm]{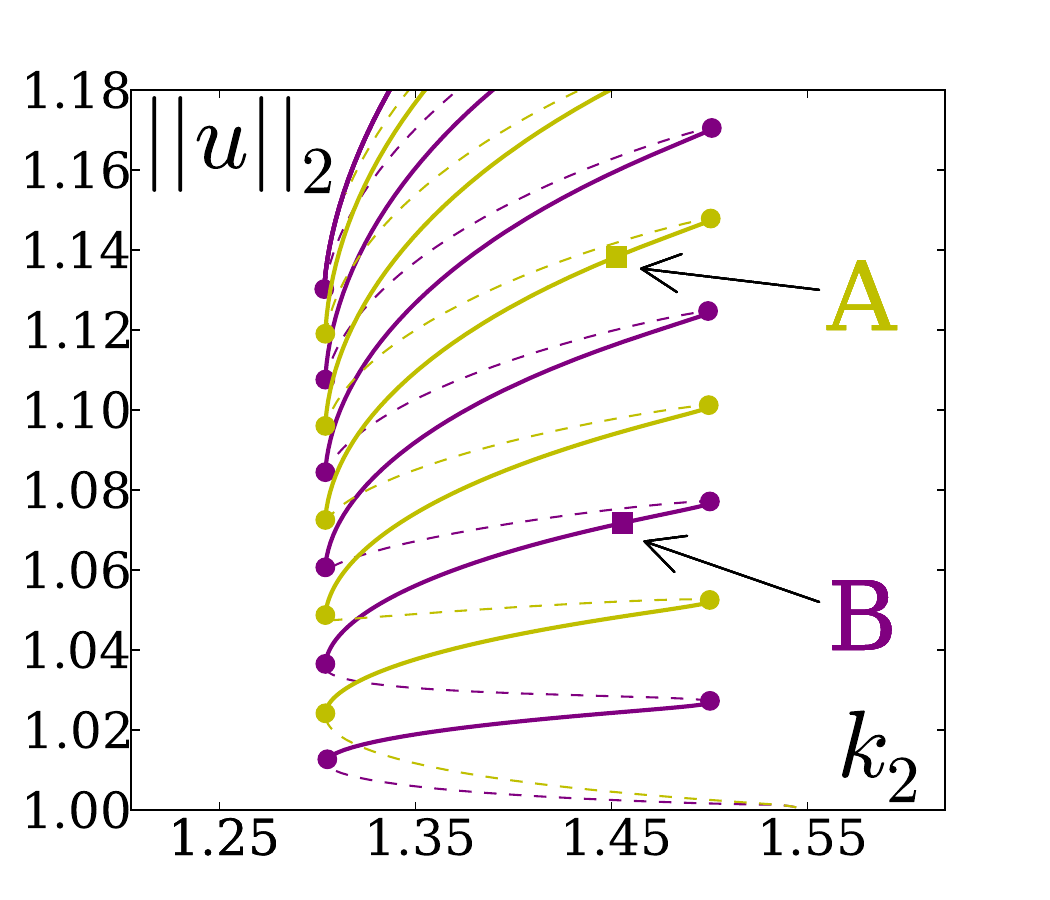}  \hspace{-0.4cm}
\includegraphics[width=4.4cm,height=4.4cm]{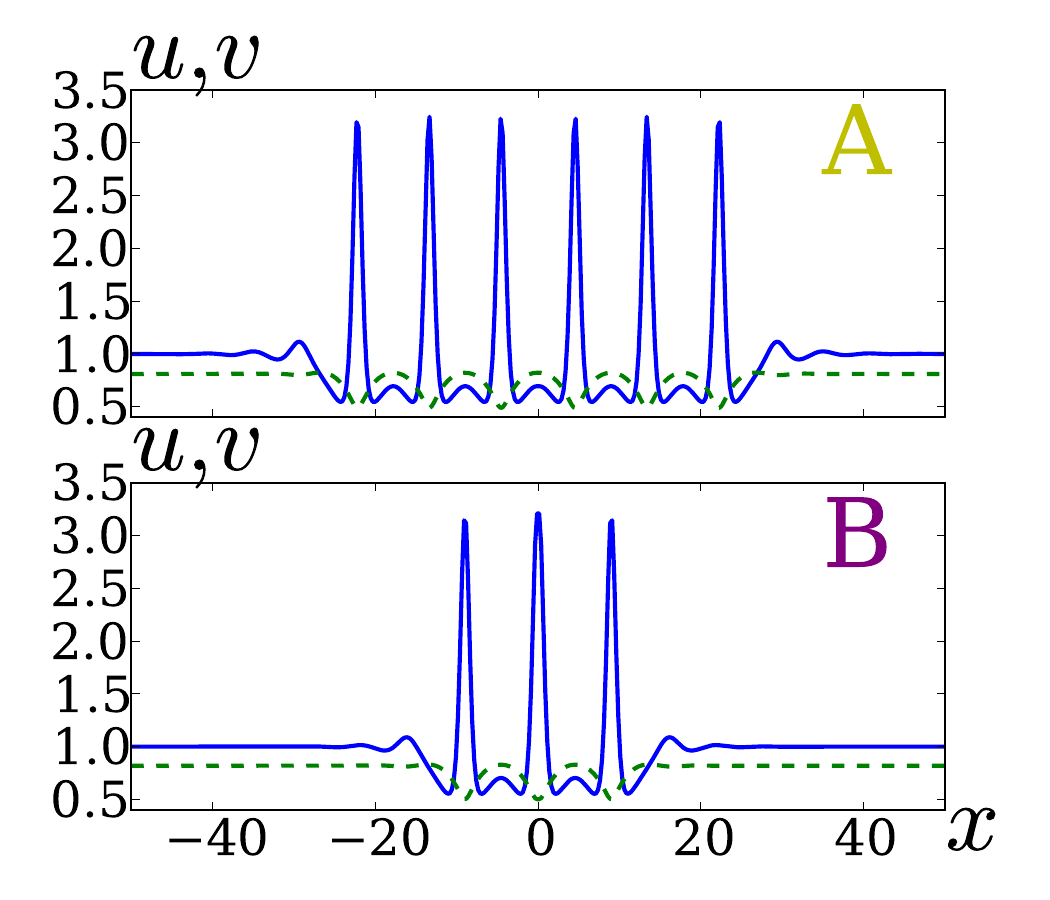}
\caption{(Color online) (a) The snaking region (shaded) inside which homoclinic
  snaking is observed in the $(k_2,D_1)$ parameter plane. (b) Homoclinic
snaking; even-solutions branch~(yellow) and odd-solutions branch
(purple), and fold bifurcations (filled circles); bold solid
lines indicate stable branches; $D_1=0.15$. (c)~Samples of
multi-pulse homoclinic stable solutions on the even- and odd-branch
for $k_2=1.45$, top and bottom panels respectively, which correspond
to 6-spike solution (label A) and 3-spike solution (label B) in~(b). The $u$-component (solid line) and
$v$-component (dashed line) are plotted for a domain size $L=100$.}
\label{fig:crittrans}
\end{center} 
\end{figure}

To calculate the criticality condition of the Turing instabilities of
\eqref{eq:fundsys}, we follow a Lyapunov--Schmidt reduction
method~\cite{golub01}. Upon obtaining the steady-state smooth
functional $\boldsymbol\phi=\boldsymbol\phi\left(U,V,\mu\right)$,
which comes from setting $u_t$ and $v_t$ to zero, and hence defining
the bifurcation function
$g\left(z,\mu\right)\equiv\left\langle\vectorr
w^*,\boldsymbol\phi\left(z\vectorr w,\mu\right)\right\rangle$ at the
steady-state $\vectorr w=\left(u-U_0,v-V_0\right)^T$, the result is
a so-called {\em bifurcation equation} $g=0$, the 
leading-order expansion of which can be written
\begin{gather}
g(z,\mu) = q_1 \mu z + q_3 z^3 + q_5z^5 +\mathcal O\left(\mu^2 z,\mu z^3\right)\, .
\label{eq:amplitude}
\end{gather}
Note that the form of $g$ is identical to the right-hand side 
of the amplitude equation \eqref{eq:long}.
The bifurcation parameter here is defined as $\mu= k_2-{k_2}_c$ 
and the scalar variable $z$ parametrises
the amplitude of the component of the Turing pattern in the kernel of
the matrix $\vectorr A_\kappa\left(\mu\right)=-\vectorr D\kappa^2+
\vectorr A$, where $\vectorr D$ is the diffusivity matrix.  
The calculation of the coefficients $q_i$ is 
straightforward but lengthy, full details are
available in \cite{brenaThesis}, we omit the details for
brevity. 

Note that such bifurcation equation \eqref{eq:amplitude} can be
derived in principle in a higher dimensional spatial domain $\Omega$.
There, a vector reduced bifurcation function~$\mathbf{g}$ can be
computed by projecting onto the eigenspace defined by modes satisfying
the boundary value problem
\begin{gather*}
\nabla^2\vectorr w+|\boldsymbol\kappa|^2\vectorr w=\vectorr0\,, \quad \left.\left(\vectorr n\cdot\nabla\right)\vectorr w\right|_{\partial\Omega}=\vectorr0\,.
\end{gather*} 

In computing the scalar function $g$, we find that trivially 
$g_{\mu}(0,0)=0$ and $g_{\mu\mu}(0,0)=0$, by virtue of the equilibrium 
being at the origin and having a zero eigenvalue. Also, owing to the reflection
symmetry in~$x$, the reduced bifurcation function must be odd in~$z$,
despite the presence of quadratic terms in the original equation.

In the parameter region under investigation the sub/super-criticality
of the pitchfork (Hamiltonian--Hopf) bifurcation is determined by the
sign of~
$$
\sigma = q_1 q_3,
$$
see \cite{iooss}. 
Figure~\ref{fig:Turdisper}(b) plots~$\sigma$, as a function of $k_2$
close to~${k_2}_c$.
We have
also checked that~$q_1q_5$ is negative in all the entire parameter
region of interest. 
The figure also shows computed bifurcating branches close
to the point where $\sigma$ changes sign, where we can see a
re-stabilising fold (limit point) in the case of the sub-critical
bifurcation. 

Now consider variation of a second parameter. For convenience we
choose $D_1$. Figure \ref{fig:crittrans}(a) depicts a two-parameter bifurcation
diagram showing the Hamiltonian--Hopf bifurcation curve, the codimension-two
point at which $\sigma=0$ and the numerically computed ``snaking region'' in
which localised states exist in the $(k_2,D_1)$ plane. 
In accordance with the usual analysis of homoclinic snaking, 
inside this region there are two branches of localised states. The states are
all invariant under the reversibility, and at each fold the
number of pulses varies, so that each second successive horizontal-like
branch has two additional large pulses. We remark that, in contrast to the
Swift--Hohenberg equation for example, there is no variational
structure in the system~\eqref{eq:fundsys}, which therefore implies there are no
asymmetric stationary localised states (so-called ``ladders'' in a
``snakes and ladders'' bifurcation diagram).

In addition, we have computed stability of the states shown in
Fig.~\ref{fig:crittrans}(b) using a standard three-point uniform
finite differences method and an eigenvalue solver.  We have
numerically found that stable branches occur similarly to results
previously found for the Swift--Hohenberg system (e.g.~\cite{burke02,Houghton})
as is shown (solid lines) in Fig.~\ref{fig:crittrans}(b). There stable
branches lose stability in a fold bifurcation (filled circles)
where branches of solutions with odd and even numbers of spikes
annihilate each other. Examples of stable solutions are shown in
Fig.~\ref{fig:crittrans}(c); note that number of spikes correspond to ladder step.

\section{2D simulation results}

\begin{figure}
\begin{center}
(a) \hspace{3.5cm}  (b) \\
\includegraphics[width=4.3cm,height=4.3cm]{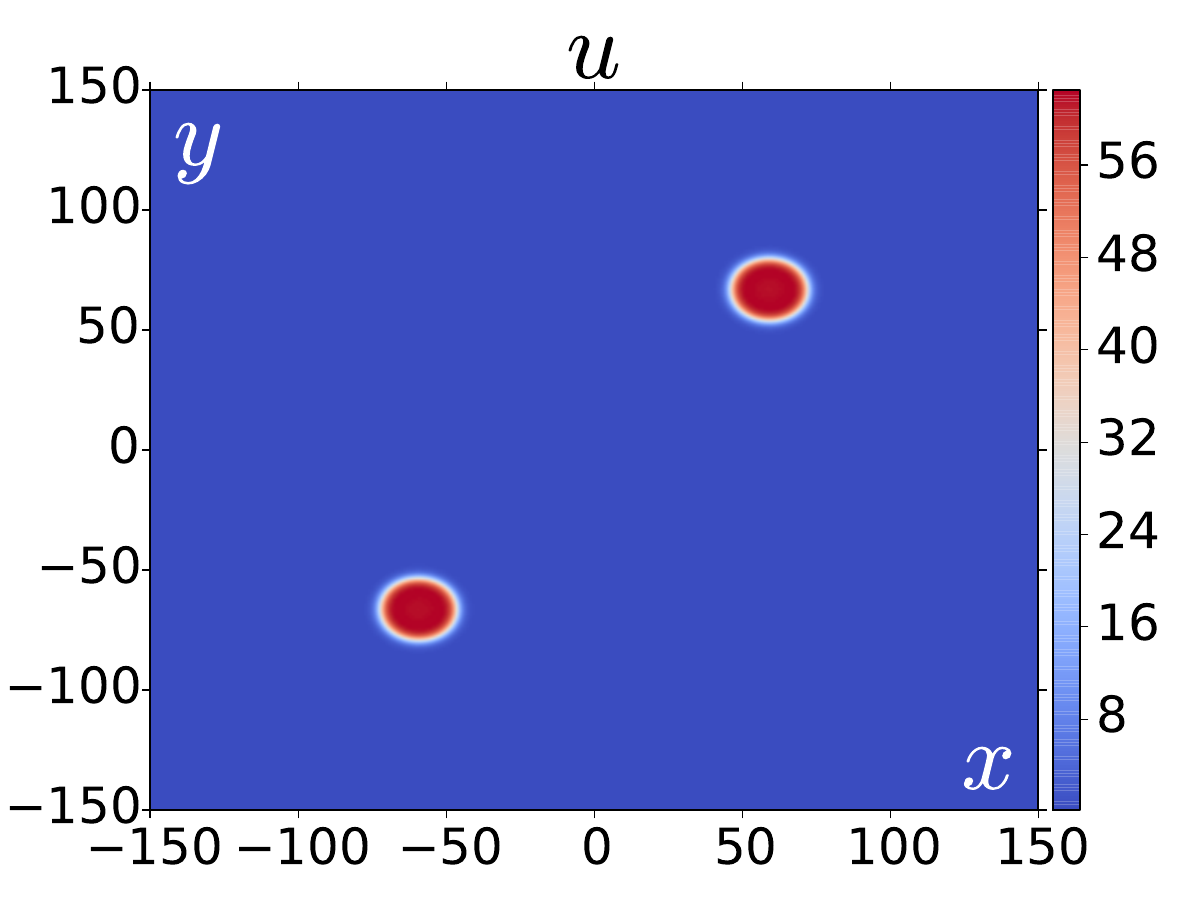}  \hspace{-0.2cm}
\includegraphics[width=4.3cm,height=4.3cm]{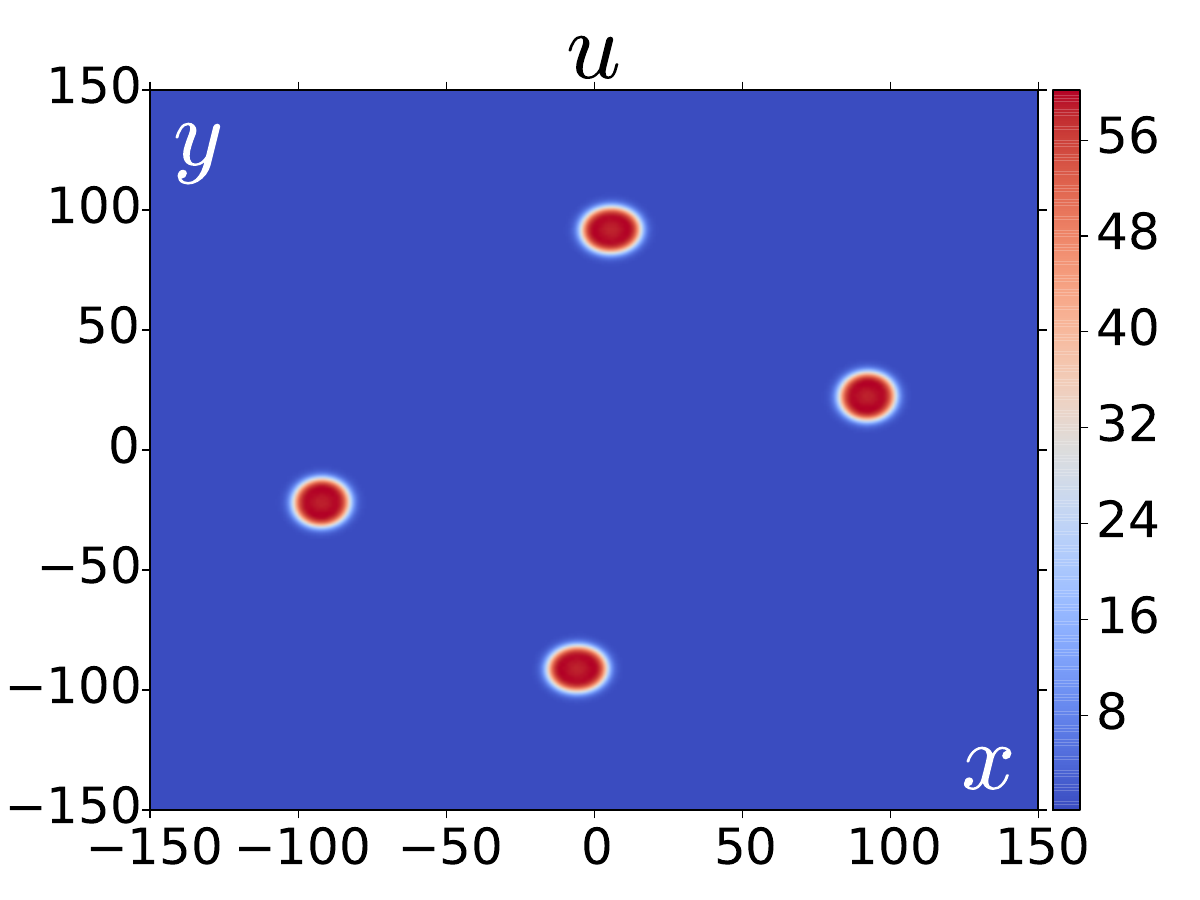} \\
(c) \hspace{3.5cm}  (d) \\
\includegraphics[width=4.3cm,height=4.3cm]{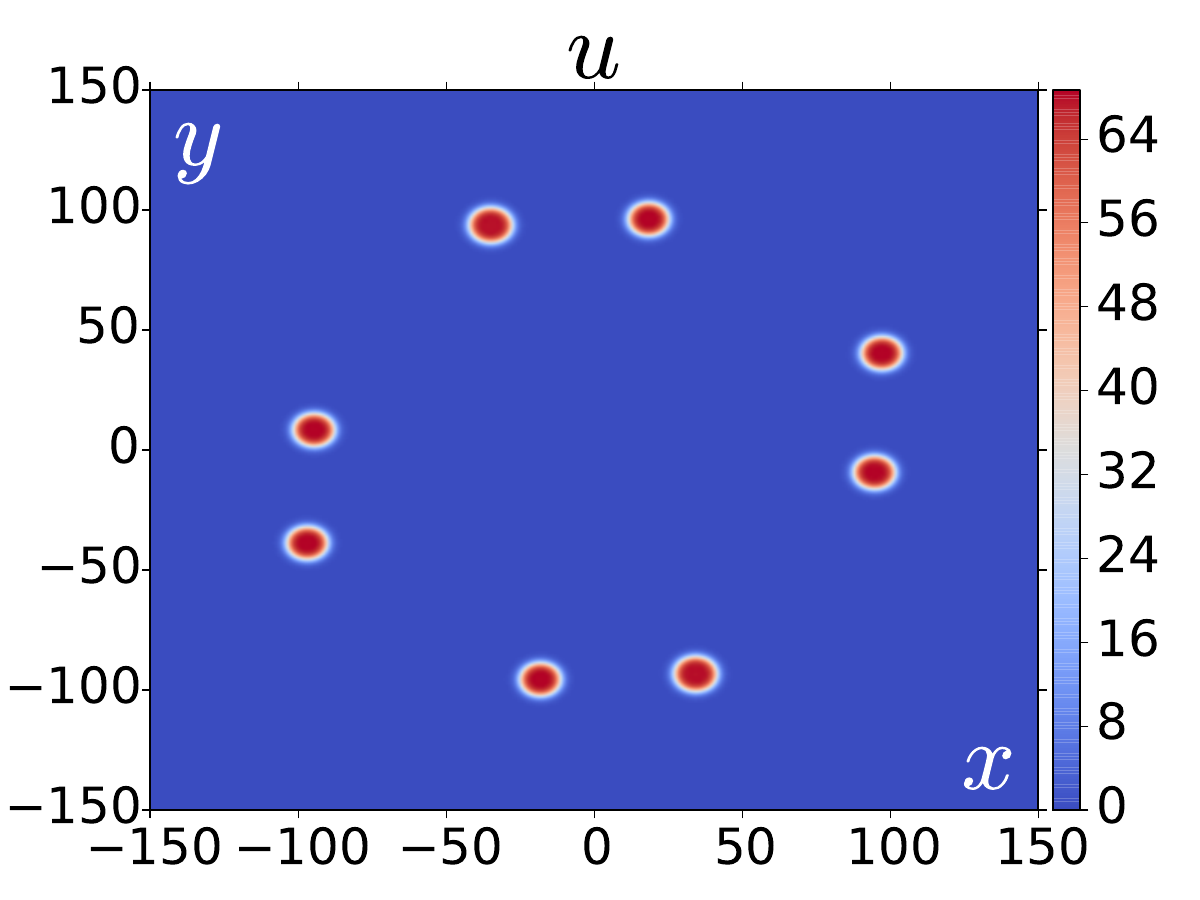} \hspace{-0.2cm}
\includegraphics[width=4.3cm,height=4.3cm]{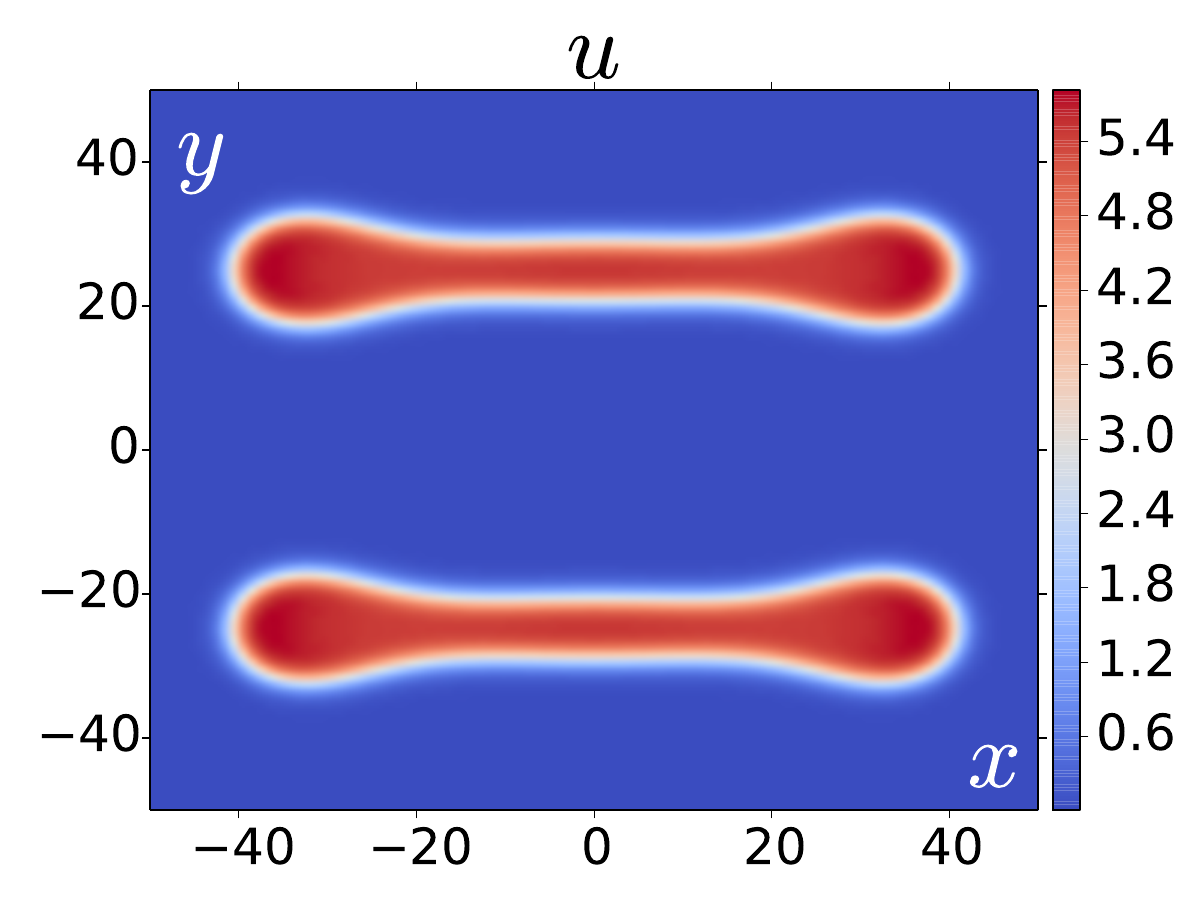} 
\end{center}
\caption{(Color online) Stable patterns for a two-dimensional square domain
  $L_{x,y}=300$ on a cool-warm scale (side bar). Localised (a)~2-spot for $r=b=10^{-4}$,
  $k_2=4\times10^{-3}$ and $D_1=0.13$, (b)~4-spot for
  $r=b=5\times10^{-4}$, (c)~8-spot for $D_1=0.11$, and other parameter
  values as in (a). (d)~2-stripe: $r=b=10^{-4}$, $k_2=0.6$, $D_1=0.1$,
  and $L_{x,y}=100$. Other parameter values: $c=0.1$, $h=10^{-2}$ and
  $D_2=10$.}
\label{fig:snak_samples}
\end{figure}

To illustrate that a similar localised pattern formation 
mechanism is likely to apply in higher spatial dimensions
we have performed simulations of the same system \eqref{eq:fundsys}, 
on a large square domain with $L_{x,y}=300$. Specifically we have used 
a finite difference method implemented in 
{\tt matlab}, with spatial resolution $300\times300$. The computations
were run a long time until steady state reached. The results are presented in 
Fig.~\ref{fig:snak_samples}.

We have found a collection of different 
localised patterns under the conditions of equal source and loss
terms $b=r$. Samples of
these can be seen in Fig.~\ref{fig:snak_samples}(a)-(c). Taking similar
initial conditions that was a $10^{-2}$ perturbation from the 
homogeneous equilibrium \eqref{eq:u0v0} resulted in the two-spot
pattern depicted in Fig.~\ref{fig:snak_samples}(a). Taking this 
solution as an initial conditions, a slight 
increase in $b$ and $r$ resulted in the 
localised 4-spot pattern  depicted in
Fig.~\ref{fig:snak_samples}(b). The dynamics of this process was
such that each spot arises through a form of 
spot-splitting dynamics \cite{nishi02}. 
In a similar fashion, we slightly 
decreased $D_1$ instead. In so doing, a localised
8-spot pattern is obtained from the 4-spot pattern, see
Fig.~\ref{fig:snak_samples}(c).

In addition, we also noted that upon performing a similar computation but
changing $b=r$ or $D_1$ in the other directions results in a either
a completely different form of spatially extended pattern or no pattern at
all. 
This suggests that hysteretic behaviour is taking place,
which should be a consequence of an overlapping structure of stable
branches similar to Fig.~\ref{fig:crittrans}(b). 


These patterns,
however, are just a few examples of the spot-like patterns that we were
able to find and indicate that the mechanism we have identified 
is likely to be robust in higher dimensions. 
A full exploration in the spirit
of~\cite{Daniele,lloyd1,lloyd2} though is left for future work.
It is interesting to note though that we were unable to find any 
localised patterns for $b \neq r$. This suggests 
that well-balanced source and loss
plays an important role in the (non-variational) pattern formation 
mechanism under investigation in higher dimensions. 

On the other hand, in Fig.~\ref{fig:snak_samples}(d), a stripe-like
pattern is depicted. We initially obtained a two-spot pattern as
before in a smaller domain, and it was taken as an initial condition
for a second run. In so doing, note that upon significantly increasing
$k_2$ and slightly decreasing $D_1$ parameter values the two-spot
pattern dramatically changes. Moreover, upon taking this solution as
an initial condition once again but in a much larger domain, a pattern
consisting of spots and stripes emerges (not shown). Patterns as such
have been observed in non-homogeneous domains,
see~\cite{paper2,plaza01}. A further analysis is nevertheless also
left for future work.

\section{Conclusions}

In summary, we have shown that canonical reaction-diffusion systems
can generate localised patterns spontaneously, which, close to a
relevant codimension-two point, can occur at small amplitude. In
particular, taking realistic source and loss terms into account can
provide conditions for a subcritical Turing bifurcation to occur. Taking
the paradigm of spatial dynamics, on a long domain, the Turing bifurcation corresponds to where two complex eigenvalues 
collide on the imaginary axis (a Hamiltonian--Hopf bifurcation). In the presence
of spatial reversibility, this provides exactly the right ingredients for
the spontaneous formation of localised patterns through the 
so-called homoclinic
snaking mechanism. Note that any realistic physical, chemical or biological
systems,
especially those found in nature, are likely to feature such source
and loss terms. 

Moreover, as argued more carefully in 
\cite{paper1} this method of pattern formation is likely to be more robust
than via the more traditional
supercritical Turing instability, and hence to be chosen by nature. 
In the supercritical case, the pattern
is always spatially extended, is born from zero amplitude and, in long
domains is typically subject to further instabilities through mode
interactions.  In contrast for the subcritical Turing bifurcations investigated
here, we find jumps to well-formed, finite amplitude patterns that are 
localised in the spatial domain. 
Due to the hysteretic nature of fold bifurcations seen in 
the snaking bifurcation branch, small fluctuations in parameter values 
or small stochastic perturbations of the kinetics would not typically 
destroy the localised pattern. 
In addition, the presence of a weak spatial inhomogeneity 
will result in bifurcation diagram similar to
that in Fig.~\ref{fig:crittrans}(b), but slanted so that the folds of
the snake occur at different values---results not shown. 
For example~\cite{paper1} 
considers the system~\eqref{eq:fundsys}, in a different parameter regime, 
in a long (but finite) domain in 1D with a spatial gradient 
multiplying the bifurcation parameter $k_2$. The result is reminiscent of
a finite portion of a slanted snake bifurcation diagram where the patterns
with a higher number of spots occur for higher $k_2$-values. This can be explained using the concept of {\em slanted snaking} established by Dawes~\cite{dawes}, where in this case the spatial gradient is replaced by a 
scalar field that is neutrally stable.  
We therefore believe the mechanism we have uncovered
will prove important in explaining and predicting observations of
localised patterns in certain a wide variety of physical systems, 
see for example~\cite{vanag}, and will only be accentuated by spatial 
homogeneity.



\end{document}